\documentclass[pre,amsmath,amssymb,aps,twocolumn, 10pt]{revtex4-2}

\usepackage{graphicx}
\usepackage{xcolor}
\usepackage{bm}
\usepackage{amsthm}
\usepackage{stmaryrd}
\usepackage{bbold}
\usepackage{hyperref}
\usepackage{float}
\usepackage{enumitem}
\usepackage{subcaption}
\usepackage{tikz}
\hypersetup{
  colorlinks=true,
  linkcolor=blue,
  filecolor=blue,      
  urlcolor=blue,
  citecolor=blue
}
\usepackage{xr}

\newcommand{\1}{\mathbb{1}}

\usepackage{etoolbox}

\DeclareRobustCommand{\parrow}{%
  \mathrel{%
    \begin{tikzpicture}[baseline=(char.base)]
      \node[draw, circle, inner sep=0.1pt, line width=0.3pt] (char) {$\to$};
    \end{tikzpicture}%
  }%
}

\DeclareRobustCommand{\marrow}{%
  \mathrel{%
    \begin{tikzpicture}[baseline=(char.base)]
      \node[draw, circle, inner sep=0.1pt, line width=0.3pt] (char) {$\gets$};
    \end{tikzpicture}%
  }%
}
\DeclareRobustCommand{\zarrow}{%
  \mathrel{%
    \begin{tikzpicture}[baseline=(char.base)]
      \node[draw, circle, inner sep=0.3pt, line width=0.3pt] (char) {$0$};
    \end{tikzpicture}%
  }%
}

\DeclareRobustCommand{\parrowf}{%
  \mathrel{%
    \begin{tikzpicture}[baseline=(char.base)]
      \node[draw, circle, inner sep=0.1pt, fill=white, line width=0.8pt] (char) {\Large$\to$};
    \end{tikzpicture}%
  }%
}

\DeclareRobustCommand{\marrowf}{%
  \mathrel{%
    \begin{tikzpicture}[baseline=(char.base)]
      \node[draw, circle, inner sep=0.1pt, fill=white, line width=0.8pt] (char) {\Large$\gets$};
    \end{tikzpicture}%
  }%
}

\usetikzlibrary{arrows.meta, positioning}

\begin{document}

\title{Activity-driven clustering and many-body steady state of jamming run-and-tumble particles}

\author{Leo Hahn$^{1,2}$}
\email{leo.hahn@unine.ch}
\author{Arnaud Guillin$^1$}
\email{arnaud.guillin@uca.fr}
\author{Manon Michel$^1$}
\email{manon.michel@uca.fr}
\affiliation{$^1$Laboratoire de Math\'ematiques Blaise Pascal UMR 6620, CNRS, Universit\'e Clermont-Auvergne, Aubi\`ere, France.}
\affiliation{$^2$Institut de Mathématiques, Université de Neuchâtel, Switzerland}

\begin{abstract}
  We exactly resolve the three-particle steady state of run-and-tumble
  particles with jamming interactions, providing the first microscopic
  description beyond two bodies. The invariant measure, derived via a
  piecewise-deterministic Markov process description and symmetry
  principles, reveals persistent, separated, and diffusive regimes
  ruled by the activity parameter. A geometric cascade of scales in
  the activity parameter organizes the structural weights, showing the
  separated phase dominates at finite activity, while non-uniformity
  plays only a minor role.  Extending these results to larger systems,
  we show that the $N$-body steady state inherits the same
  organization: the number of clusters becomes sharply defined by the
  activity value, with crossover boundaries whose slopes diverge with
  $N$. We also show how the activity plays a role similar to a
  fugacity conjugate to cluster number, yielding a
  grand-canonical–like structure emerging directly from the
  microscopic dynamics. This framework lays the groundwork for a
  systematic microscopic theory of active many-body steady states.
\end{abstract}

\maketitle

\definecolor{myGreen}{rgb}{0, 0.4, 0}
\definecolor{qqqqff}{rgb}{0,0,1}
\definecolor{uququq}{rgb}{0.25,0.25,0.25}

\newcommand{\red}[1]{{\color{red} {\bf Caution ! } #1}}
\newcommand{\vertex}{{\mathcal J}}
\newcommand{\edge}{{\mathcal S}}
\newcommand{\Fvertex}{{\bar{\mathcal J}}}
\newcommand{\Fedge}{{\bar{\mathcal S}}}
\newcommand{\edgeExp}{_\mathcal{S}^\mathrm{rel}}
\newcommand{\edgeUnif}{_\mathcal{S}^\mathrm{eq}}
\newcommand{\bulk}{{\mathcal B}}
\newcommand{\unif}{{\mathrm{eq}}}
\newcommand{\rel}{{\mathrm{rel}}}
\newcommand{\Fbulk}{{\bar{\mathcal B}}}

\begin{figure*}
	\includegraphics[width=\textwidth]{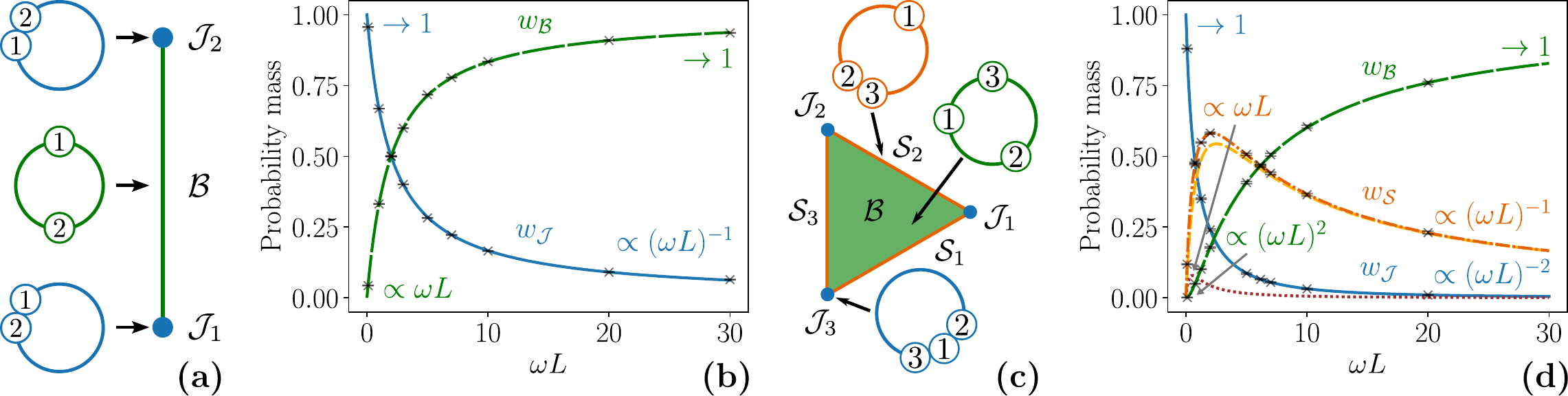}
	\caption{(a,c) State space for the 2-RTP (Bulk and Jammed
          states) and 3-RTP (Bulk, Separated and Jammed states)
          systems respectively. (b,d) Weights in the steady state
          (jammed $w_\vertex$ (green, long dashes), bulk $w_\bulk$
          (blue, solid) and separated $w_\edge$ (orange, dash-dotted))
          as a function of $\omega L$ for the 2-RTP and 3-RTP systems
          respectively. Grey cross markers denote simulation results
          and exhibit excellent agreement (see Appendix~\ref{sec:simu}
          for details). For the 3-RTP system, $w_\edge$ decomposes
          into its uniform $w_\mathrm{eq}w_\edge$ (yellow, dashed) and
          exponential part $w_\mathrm{rel}w_\edge$ (maroon, dotted);
          $w_\edge$ achieves a maximum of $\approx 0.582$ at
          $\omega L \approx 2.16$.  For
          $\omega L \le 3w_\unif/(5-2w_\unif)\approx 0.759$, jammed
          states dominate, then for
          $0.759\approx \le \omega L \le 6/w_\unif\approx 6.16 $
          separated ones do, after which the free states dominate. In
          the 2-RTP case, $w_\vertex=w_\bulk$ at
          $\omega L = 2$.}\label{fig:main_figure}
\end{figure*}

{\noindent \bf Introduction.} Active matter, ensembles of units that
convert stored or ambient energy into directed motion, has emerged as
a cornerstone in nonequilibrium statistical physics, now spanning
scales from cytoskeletal filaments, pedestrian crowds to robotic
swarms \cite{schnitzer93, wu00,
  ramaswamy10,cates12,marchetti13,dauchot19,mayya19, wang21}. Within
this field, {\it run-and-tumble particles} (RTPs) furnish a
fundamental microscopic model
\cite{schnitzer93,berg04,kafri08,tailleur08,slowman16,slowman17,malakar18,ledoussal19,das20,basu20,mori20,metson22,arnoulx23,hahn23,guillin24}:
persistent ballistic runs punctuated by Poissonian velocity tumbles
break detailed balance and give rise to hallmarks of active behavior
such as wall accumulation~\cite{elgeti15}, coarsening
\cite{soto14,sepulveda16} and motility-induced phase separation (MIPS)
\cite{tailleur08,fily12,marchetti13}.

In the presence of hard-core repulsion, RTPs indeed display phenomena
qualitatively different from their passive
counterparts. Activity-driven {\it jamming} interactions, clustering
\cite{fily12,soto14,sepulveda16,slowman16,slowman17} and different critical
behaviours \cite{bi16} emerge. Clustering is often rationalized in
terms of an effective attraction: in dense regions particles slow
down, further increasing density and triggering
aggregation~\cite{cates15,dolai20}. Yet this picture is essentially
coarse-grained~\cite{toner95,tailleur08,bialke13,marchetti13,cates15,farage15},
and exact microscopic results have remained elusive. Beyond two
particles, where the problem reduces to a single effective RTP with
more complex velocity dynamics~\cite{slowman16,das20,ledoussal21}, the
steady state remains unknown. Despite an important theoretical
interest, the absence of exact three-body or many-body solutions has
been a major barrier to a first-principles understanding of clustering
and to clarifying the microscopic origin of coarsening or MIPS phenomena.

Indeed, a major challenge lies in handling the singular boundary
conditions introduced by jamming~\cite{angelani17,bressloff22}, even
in the two-RTP case. Lattice
discretizations~\cite{slowman16,metson22}, soft repulsion
limits~\cite{arnoulx23}, or added noise~\cite{malakar18,das20}
regularize the problem but at the cost of complicating the original
process and adding limits nontrivial to compute
\cite{guillin24}. Progress came with the use of piecewise
deterministic Markov processes (PDMPs)~\cite{hahn23}, whose generators
encode both deterministic runs and stochastic tumbles, including
boundary behavior, directly in continuous time and space. For two
RTPs, this framework yielded the explicit steady state for arbitrary
tumble mechanisms and thus revealed the existence of {\it detailed-}
and {\it global-jamming} system classes. These classes are linked to a
dynamical symmetry, whose breaking creates relaxation terms in the
steady state, that are often interpreted as effective attractions. Yet
despite this progress, key questions remain open regarding
activity-driven clustering beyond two bodies and it is still unclear
to what extent advances in the exactly solvable two-particle case can
inform this broader problem. Can the mechanisms uncovered for pairs,
such as dynamical-symmetry breaking and the resulting effective
attractions, survive in higher-body encounters, or do fundamentally
new behaviors emerge? Moreover, simulations consistently report
  robust clustering patterns \cite{soto14,sepulveda16}, but these
  observations still lack a microscopic analytical
  derivation. Establishing exact steady states is therefore crucial:
  not only would they clarify the origin of these collective
  structures, but they would also provide much-needed benchmarks for
  testing and validating numerical schemes.

In this paper, we resolve the three-body
invariant measure, provide the first controlled microscopic
description of clustering beyond two bodies and establish a
structural foundation for the $N$-body problem:
\begin{itemize}[leftmargin=*,label=\raisebox{0.25ex}{\tiny$\bullet$}]
  \item Extending the PDMP
and symmetry formalism developed for two RTPs~\cite{hahn23}, we derive
the invariant measure of three instantaneous RTPs. This measure
identifies with a mixture over free, separated, and jammed states. The
relative strength of the structural weights setting this mixture
organize persistent, separated, and diffusive regimes through the
activity parameter $\omega L$, as displayed in
Fig.~\ref{fig:main_figure}.
\item Crucially, the hierarchy of structural weights follows a geometric
cascade, $(\omega L)^{n_C}$, with $n_C$ the number of clusters,
revealing that the separated phase dominates at finite
activity. Relaxation contributions, previously interpreted as
effective attractions, are subdominant and do not drive
clustering.
\item Extending these insights to large $N$, we find that the
cluster number becomes sharply defined for a given $\omega L$, with
crossover boundaries whose slopes diverge as $N$ grows. In this sense,
we discuss how the activity parameter plays a role analogous to a
fugacity conjugate to cluster number, giving rise to a nonequilibrium
grand-canonical-like structure that emerges directly from the
microscopic dynamics.\end{itemize}

Therefore, these results not only provide a principled microscopic
  foundation for many-body steady states in active systems, but also
  highlight the subtle interplay between persistent motion, jamming,
  and activity in shaping cluster formation and stability. By bridging
  the gap in exact solutions beyond two particles, our work offers
  clear guidance for numerical and experimental studies targeting the
  parameter regimes of clustering and phase separation of
  interest. Importantly, the structural insights and hierarchical
  organization we uncover provide concrete directions for tackling the
  major challenge of resolving the full combinatorial structure
  of the $N$-body steady state. Achieving this would enable a complete
  understanding of phase separation and a rigorous microscopic
  characterization of coarsening and MIPS, opening a promising path
  for future analytical progress.\medskip 

In the following, we begin by recalling the recent results on symmetry
constraints and steady-state structure in the two-RTP case, as these
tools are essential for deriving the invariant measure of three
interacting particles. Building on this foundation, we then construct
the full three-RTP steady state by combining symmetry arguments with
the PDMP framework, which provides an efficient way to handle the
deterministic motion and boundary-induced transitions characteristic
of jamming. With the explicit three-body measure in hand, we analyze
how phase separation is stabilized at finite activity, identifying the
regimes governed by structural weights and the geometric scales they
generate. Finally, we show how most of these results extend to
arbitrary particle number $N$, unveiling the hierarchical organization
of clusters and the emergence of a sharply defined cluster number in
the large-$N$ limit.

\section*{Jamming RTP and dynamical symmetry}

We consider a system of RTP on a
1D torus of length $L > 0$. Each particle~$i$ moves with velocity
$\sigma_i$ (\emph{Run}) updated by an independent Markov jump process
(\emph{Tumble}). Collisions lead to jamming, where both particles
remain immobile until a velocity update allows separation.  The system $z=(r,\sigma)$
is fully specified by the inter-particle separation vector
$r=(r_i)_i$, where $r_i$ denotes the distance between particle $i$ and
particle $(i+1)$, together with the velocity vector
$\sigma=(\sigma_i)_i$. A periodic convention is used for the
indices/particle labels, i.e. $x_{i} = x_{i\pm 3}$. For two RTP, the
steady state was solved for arbitrary tumbles~\cite{hahn23} and is of
the form
$\pi = w_\bulk \pi_\bulk + \frac{w_\vertex}{2}\sum_{k=1}^2
\pi_{\vertex_k},$ where $\pi_\bulk$ is supported on free states
$\bulk$, and $\pi_{\vertex_k}$ on jammed states, $\vertex_1,\vertex_2$
(boundaries of $\bulk$, see
Fig.~\ref{fig:main_figure}), 
 \begin{align*}
   	\pi_\bulk(z) =  \frac{w_{\rm eq}}{L}\mu_{\mathcal{B}}(\sigma) + w_{\rm rel}\gamma(z), \;
   \pi_{\vertex_k}(z) =  \mu_{\vertex}(\sigma_k,\sigma_{k+1}). 
 \end{align*}
 The distributions $\mu_\bulk$ and $\mu_{\vertex}$ are the respective tumble
 steady states, left invariant by the jump process for
 $\sigma$.

\begin{figure}
  \includegraphics[width=.45\textwidth]{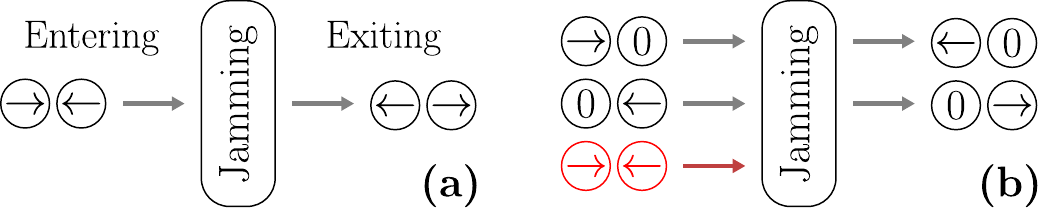}
  \caption{(a) Detailed-jamming is realized for instantaneous tumbles
    ($\marrow \leftrightarrow \parrow$). (b) Only global-jamming is achieved for finite
    tumbles
    ($\marrow\leftrightarrow \zarrow \leftrightarrow \parrow$):
    exiting with $\marrow \parrow$ is impossible as
    $\mu_\vertex(\parrow\parrow) =\mu_\vertex(\marrow\marrow)=0$.}
        \label{fig:det-glob-jam}
\end{figure}

 The crucial observation of \cite{hahn23} is that conservation of
 probability flows defines an \emph{active global balance} in
 $\bulk$. Detailed satisfaction of this balance requires
 $\pi(r,(\sigma_1,\sigma_{2})) = \pi(r,(\sigma_2,\sigma_1))$, which
 $\mu_\bulk$ automatically possesses due to particle
 indistinguishability. Thus, deviations from detailed symmetry in
 $\sigma$ arise from the imbalance between entering and exiting
 scenarios at jamming boundaries, which equivalently reflect the
 incompatibility between $\mu_\bulk$ and $\mu_\vertex$, as shown in
 Fig.~\ref{fig:det-glob-jam}. Accordingly, detailed-jamming systems
 necessarily have $w_{\text rel}=0$, maintaining a uniform steady
 state $\propto \mu_\bulk$ in $\bulk$, identical to the passive
 equilibrium distribution. This occurs for \emph{instantaneous-tumble}
 RTPs, where velocities $\sigma_i=\pm1$ flip with rate $\omega>0$. By
 contrast, global-jamming systems break this detailed symmetry. The
 system is shifted away from uniformity at the jamming boundary,
 resulting in a catenary relaxation term $w_{\text rel}\gamma$. This
 arises for \emph{finite tumble} RTPs, where $\sigma_i\in\{-1,0,+1\}$
 and $\gamma$ is a single-scale catenary relaxation.

This highlights the interplay between the richness of
tumbling mechanism, symmetry preservation and boundary-induced
relaxation. A natural next question is whether increasing the number
of particles alone can generate non-equilibrium deviations
away from uniformity. In particular, can three instantaneous RTPs
already produce a non-uniform steady state---potentially stabilizing
separated phases?\smallskip

\section*{Exact three-RTP steady state} Increasing the particle number
alone gives rise to qualitatively new phenomena: in the three-particle
system, three-particle clusters and phase-separated states emerge,
with two particles jammed and one remaining free. By deriving the
exact invariant measure for a three-RTP system with instantaneous
tumbles, we show that the symmetry picture remains relevant and
demonstrate the stabilization of phase-separated configurations.

As illustrated in Fig.~\ref{fig:main_figure} and detailed in
Appendix~\ref{app:conservation}, the state space $\Omega$ is split
into: the bulk $\bulk$ where all particles are free, the three edges
$\edge_k$ where the system is separated into a cluster of two jammed
particles and a free one and the three vertices $\vertex_k$ where all
particles are jammed into a single cluster. The steady state $\pi$ is
a mixture over distributions $\pi_\bulk,\pi_{\edge_k},\pi_{\vertex_k}$
respectively supported on the free, separated and jammed states,
 \begin{equation}
   \pi = w_\bulk \pi_\bulk + \frac{1}{3}\sum_{k=1}^3 [w_{\edge} \pi_{\edge_k} + w_{\vertex} \pi_{\vertex_k}].
   \label{eq:pi-gen}
 \end{equation}
We show that, for
$(r,\sigma)\in \bulk$ (resp. $\vertex_k$, $\edge_k$),
  \begin{equation}
    \begin{aligned} 
    \pi_\bulk(r,\sigma) &=  \frac{2}{L^2}\mu_\bulk(\sigma),\\
      \pi_{\edge_k}(r,\sigma)&= \frac{w_{\rm eq}}{L}\mu_{\edge}(\tau_k(\sigma))  + w_{\rm rel}\gamma(r_{k+1},\tau_k(\sigma)),\\
      \pi_{\vertex_k}(r,\sigma) &=  \mu_{\vertex}(\tau_k(\sigma)),
  \end{aligned}
  \label{eq:pi}
\end{equation}
where, for any vector x, the
$\tau_k(x_1,x_2,x_3)=(x_k,x_{k+1},x_{k+2})$ are the index shifts. The
distributions $\mu_\bulk$, $\mu_\edge$ and $\mu_\vertex$ stand for the respective
tumble steady states. There is no relaxation in the free
states and the relaxation term $\gamma$ in the separated ones takes
the form,
\begin{multline}
  \gamma(r,\sigma)=
  \frac{1}{\mathcal{N}_\gamma}\Big[\nu_S(\sigma)\;{\rm ch}\Big(\lambda \Big(\frac{L}{2}-r\Big)\Big)
\\  + \sigma_{3}\nu_A(\sigma)\;{\rm sh}\Big(\lambda\Big(\frac{L}{2}-r\Big)\Big)\Big]
  \label{eq:gamma}
  \end{multline}
  where $\lambda = 2\sqrt{2} \omega$ and
  $\mathcal{N}_\gamma=\frac{16{\rm sh}(\lambda L/2)}{\lambda}$.
  Appendix~\ref{app:full-pi} summarizes the full formula. At
  the marginal level, the state space reduces to the simplex set by $r_1+r_2+r_3 = L$ and $\pi$
  sums up to a pure catenary relaxation along each
  separated-state edge,
\begin{multline}\label{eq:r_marginal}
 \pi(r) = w_\bulk\frac{2 \, \1_\bulk(r)}{L^2} + \frac{1}{3} \sum_{k = 1}^3 \Big[ w_{\vertex}\1_{\vertex_k}(r) \\
  + w_{\edge} \1_{\edge_k}(r)\Big(\frac{w_\unif }{L} + w_\rel \frac{\lambda {\rm ch}\left(\lambda( L/2-r_{k+1})\right)}{2 {\rm sh}(\lambda L/2)}\Big)\Big]. 
\end{multline}
Compared to the 2-RTP case, the 3-RTP steady state still reflects
particle indistinguishability, appearing as a sum structure, and
remains uniform over free states.  The crucial novelty is the
emergence of phase-separated configurations, where the steady state is
not of product form, marking a stronger departure from equilibrium
driven purely by the increase in particle number.  In these states,
relaxation proceeds in a catenary fashion under the constraint imposed
by jamming boundaries.  This reflects a fundamental change in
symmetry: while instantaneous tumbles enforce detailed-jamming
symmetry at the free/separated boundary (i.e. (un)jamming of a pair of
particles), the jamming pair within separated states behaves as an
effective single particle with richer tumble dynamics---akin to a
finite tumble with a non-motile state---thereby breaking
detailed-jamming and producing global-jamming relaxation.

The central question of the three-RTP problem is the stabilization of
this separated phase. To address this, we show how we obtained the
explicit steady state using PDMP, including the weights (see~
\eqref{eq:nu}, \eqref{eq:muj}, \eqref{eq:wS_wB_wJ}). For further details
on the PDMP approach for RTPs, see
\cite{hahn23,guillin24}.  Here, the PDMP is
characterized by its generator, for any state $z=(r,\sigma)$ and $f$
some test function,
\begin{equation} \label{eq:generator}
\mathcal Af(z) = \underbrace{\left \langle \phi(z), \nabla_r f(z)\right\rangle}_\text{Run}+ \omega\sum_{j}\underbrace{\left(f(\iota_j(z)) - f(z)\right)}_\text{Tumble},
\end{equation}
where each $\iota_k$ flips the $k$-th particle velocity,
$\sigma_k \to -\sigma_k$, translating the $k$-th particle tumble at rate
$\omega$, and the flow $\phi$ encodes the dynamics of $r$ set by the
velocities $\sigma$, with
$\dot r = \phi(z) = (\phi_i(z))_i$, with,
\begin{equation}\label{eq:phi}
   \phi_i(z)\!=\! \!\left\{\!  \begin{array}{l}
      0 \ \text{if} \  (r_i, \sigma_{i+1} \!-\! \sigma_i)\!\!\in\! \{(0,-2),(0,0),(L,2)\}\! \\
                         \!-\!\sigma_i \ \text{if} \ (r_{i+1}, \sigma_{i+2} \!-\! \sigma_{i+1})\!=\!(0,-2), r_i > 0 \\
                         \sigma_{i+1} \ \text{if} \ (r_{i+2}, \sigma_{i} \!-\! \sigma_{i+2})\!=\!(0,-2), r_i > 0 \\
                         \sigma_{i+1} \!-\! \sigma_i \ \text{otherwise}.
             \end{array}\right.
         \end{equation}         
         A key feature of the dynamics is the emergence of velocities
         $\phi_i(z) = \pm 1$ in separated states down to $\phi=0$ in
         jammed ones, which underlies the velocity slowdown as
         clustering and separation develop.

         As detailed in
         Appendix~\ref{app:conservation}, the invariance condition
         $\int_{\Omega} \mathcal Af\, d\pi = 0$ translates into the
         following conservation laws on probability flows, with
         contributions ordered as tumble, run, and boundary
         exchanges (the latter arising from integration by
         parts of the run term in \eqref{eq:generator}),
               {\setlength{\abovedisplayskip}{2pt}
\setlength{\belowdisplayskip}{2pt}
         \begin{multline}\tag{$\textrm{C}_{\bulk}$}\label{eq:cb}
           \sum_{j} \underbrace{(\pi_\bulk(\iota_j(z)) -  \pi_\bulk(z))}_{\text{Tumble}}   =     \sum_j \underbrace{\frac{\phi_j(z)}{\omega}\partial_{r_j}\pi_\bulk(z)}_{\text{Run}}  + \underbrace{0}_{\text{Bdry}}
         \end{multline}
         \begin{multline}
  \sum_{j}\underbrace{(\pi_{\edge_k}(\iota_j(z)) -  \pi_{\edge_k}(z))}_{\text{Tumble}}   =\sum_j \underbrace{\frac{\phi_{j}(z)}{\omega}\partial_{r_{j}}\pi_{\edge_k}(z)}_{\text{Run}} \\ + \underbrace{\frac{3 w_\bulk}{w_\edge}\1_{\partial_k \bulk}(z)\lim_{\zeta\in\bulk\to z}\frac{\phi_k(\zeta)}{\omega} \pi_\bulk(\zeta)}_{\text{Boundary}} \tag{$\textrm{C}_{\edge_k}$} \label{eq:cs}
\end{multline}
\begin{multline}
  \sum_{j}\underbrace{(\pi_{\vertex_k}(\iota_j(z))-\pi_{\vertex_k}(z) )}_{\text{Tumble}}
  =   \underbrace{0  }_{\text{Run}}
  \\\underbrace{- \frac{w_\edge}{w_\vertex}\sum_{k'}\1_{\partial_k\edge_{k'}}(z)\lim_{\zeta\in\edge_{k'}\to z}\frac{\phi_{k+2}(\zeta)}{\omega}\pi_{\edge_{k'}}(\zeta)}_{\text{Boundary}} \tag{$\textrm{C}_{\vertex_k}$}
    \label{eq:cj}
         \end{multline}}   
         where $\partial_k \bulk$ and $\partial_{k} \edge_{k'}$
         respectively gather the entering and exiting boundary points
         between free/separated and separated/jammed phases, i.e.~the
         annihilation or creation of the jamming pair through particle
         transport.

         Similarly to the 2-RTP case \cite{hahn23}, the conservation
         of probability flows is ensured by the compensation between
         tumble and run terms and the exchanges on the boundary
         between the different phases. However, in addition to the
         free and jamming phases, the novel separated phase
         complexifies the condition hierarchy. Fortunately, we can
         extend the dynamical symmetry approach stemming from the
         active global balance \cite{hahn23} to restrict even
         further those conditions and provide the complete unique solution:

         \noindent (C$_\bulk$) Since tumbles are
         instantaneous, the detailed dynamical symmetry,
         (e.g.
         $P(\textcolor{red}{\parrow\marrow} \marrow) \!=\!
         P(\textcolor{green!60!black!80}{\marrow\parrow} \marrow)$),
         can be satisfied at the free/separated boundary (see
         Appendix~\ref{app:tumble-fig}). This leads to
         $\pi(r,\sigma) = \pi(r,-\sigma), (r,\sigma)\in
         \bulk$. Solving \eqref{eq:cb} under this symmetry necessarily
         cancels the derivative term and fixes the contribution
         $\pi_{\bulk}$ in \eqref{eq:pi-gen} to be the uniform solution
         of \eqref{eq:cb}, set by the bulk tumble measure,
         \begin{equation} \mu_\bulk(\sigma) = 1/8.\end{equation}

           \noindent (C$_{\edge_k}$) We refine the general
           expression of $\pi_{\edge_k}$ \eqref{eq:pi} by identifying
           $\gamma$ as the relaxation signature of broken dynamical
           symmetry at jamming. As shown in
           Appendix~\ref{app:tumble-fig}, jamming and unjamming states
           at the separated/jammed boundary cannot be balanced in a
           detailed-jamming manner
           (e.g. $P(\textcolor{red}{\parrow}\parrow\marrow) \!\neq\!
           P(\textcolor{green!60!black!80}{\marrow}\parrow\marrow)$)
           due to the incompatibility of the tumble mechanisms in the
           respective phases. To determine $\gamma$, we first exploit
           the uniform form of $\pi_\bulk$ which generates constant
           source terms in \eqref{eq:cs}. Their effect on the
           separated phase can be viewed as a modification of the
           tumble mechanism, since outflows from certain
           configurations are reinterpreted as inflows into others.
           As detailed in Appendix~\ref{app:relax}, this leads
           directly to the pure catenary relaxation,
\begin{equation}  
  \partial_{\bar{r}\bar{r}}^2\left(
\bm{\pi}^+_{\edge_k}\pm\bm{\pi}^-_{\edge_k}
   \right)
     = \lambda^2L_\pm \left(
\bm{\pi}^+_{\edge_k}\pm\bm{\pi}^+_{\edge_k}
   \right), \lambda = 2\sqrt 2\omega,
  \label{eq:2ndPDE}
\end{equation}
where
$\bm{\pi}^\pm_{\edge_k}(\bar{r})=(\pi_{\edge_k}(r,\sigma^\pm_n))_{n=1}^2$,
with $\bar{r}=r_{k+1}$ and $\sigma^\pm_n$ so that the flow
$\phi_{k+1}(r,\sigma^\pm_n) = \pm n$. The matrices $L_\pm$ admit as
eigenvalues $0$ and $1$, setting $\gamma$ to the form
\eqref{eq:gamma}. Further constraints from the eigenvectors of
$L_\pm$, symmetry around $L/2$ and first-order arguments lead to,
\begin{equation}
	\def\arraystretch{1.4}
    \begin{array}{@{}ll@{}}
        \nu_S(\sigma)\!=\!\left\{ \begin{array}{@{}l@{}}
         \sigma_1\sigma_3 \\
      4
      \end{array}\right.&
      \nu_A(\sigma)\!=\!\left\{ \begin{array}{@{}lr@{}}
      \sigma_1\sigma_3\sqrt{2}  &\ \text{if} \ \sigma_1=\sigma_2\\
      2\sqrt{2} &\ \text{otherwise} 
      \end{array} \right.     \end{array}.
  \label{eq:nu}
\end{equation}

The separated tumble measure $\mu_\edge$ is the uniform solution of
\eqref{eq:2ndPDE}, but also of \eqref{eq:cs} once composed
with $\tau_k$, setting the weight ratio between free and separated
states,
\begin{equation}  
  \mu_\edge(\sigma)=\frac{2-\delta_{\sigma_1,\sigma_2}}{8},  \quad \frac{w_\bulk}{w_\edge w_{\text{eq}}}=\frac{\omega L}{6}.
  \label{eq:mus}
\end{equation}
Therefore, at the free–separated boundary, the exchanges does not
involve the term $w_\edge w_{\rm rel}\gamma$ and reduces to a mapping of the
tumble term $w_\edge w_{\rm eq}\mu_\edge$ onto the bulk source terms,
consistent with their role as tumble modifiers.

Parallels can be drawn between $\pi_{\edge_k}$ and a generic 2-RTP
system, but with crucial differences. A separated configuration maps
onto an effective two-particle system, now distinguishable---the
single particle and the jamming pair exhibit different tumbles---and
further impacted by bulk source terms. A simple catenary relaxation
term was expected from applying spectral argument \cite{hahn23} to the
effective system but particle are now distinguishable and it manifests through the
existence of an antisymmetric ${\rm sh}$ contribution even in fully
aligned $\pm\pm\pm$ states.

  \noindent (C$_{\vertex_k}$) As shown in
  Appendix~\ref{app:jam}, the jammed constraints fix
  the jammed tumble measure $\mu_\vertex$ and determine
  $w_{\rm eq}=1-w_{\rm rel}<1$, as the detailed-jamming breaking
  generates the relaxation $w_{\rm rel}\gamma$ in the separated phase,
  \begin{equation}
    \mu_\vertex(\sigma) = \frac{3-2\delta_{\sigma_1,\sigma_3}}{8}, 
      w_\unif\!=\!  1\!-\! \frac{1}{1+ \omega L\left[\frac{2\sqrt 2}{{\rm th}( \omega L\sqrt 2)}  \!+\! \frac{10}{3}\right]}.
    \label{eq:muj}
  \end{equation}  
    \smallskip\noindent ($w$) As shown in Appendix~\ref{app:weights}, the weights
  $w_\bulk,w_\edge,w_\vertex$ are set by expliciting the boundary
  terms in \eqref{eq:cs} (i.e. \eqref{eq:mus}), and in \eqref{eq:cj}
  and the normalisation condition,
\begin{equation}\label{eq:wS_wB_wJ}
 \begin{aligned}
   w_\bulk&\!=\!\frac{(\omega L)^2w_\unif}{N_w}, \; w_\edge\!=\!\frac{6\omega L}{N_{w}}, \; w_\vertex\!=\!\frac{ ( 6\!+\!4\omega L)w_\unif \!-\!4\omega L}{N_w}\\
  N_{w}&=2\omega L+ w_\unif\left[6+4\omega L + (\omega L)^2\right].
 \end{aligned}
\end{equation}
  
\section*{Separated-phase stabilization} We now discuss the
central issue of the stabilization of the separated phase through the
weight competition, as shown in
Fig.~\ref{fig:main_figure}. The explicit weights~\eqref{eq:muj} and
\eqref{eq:wS_wB_wJ} reveal three distinct regimes, controlled solely
by the activity $\omega L$. The separated uniform component
$w_\unif$ remains $\sim 1$ throughout, tending to $2/3$ as
$\omega L \ll 1$ and to $1$ as $\omega L \gg 1$. Therefore, it is a natural
reference for reading the regimes directly from the weight hierarchy
in~\eqref{eq:wS_wB_wJ}. In the \emph{persistent} regime
$\omega L \ll 1$, the weights scale as $w_\vertex \approx 1$ and
$w_\edge \propto \omega L, w_\bulk \propto (\omega L)^2$. Therefore,
the system predominantly forms a single three-particle cluster,
spending almost all its time in fully jammed configurations. This
reflects the fact that particles traverse the state space much faster
than they tumble, yielding effective ballistic dynamics. In the
opposite \emph{diffusive} regime $\omega L \gg 1$, the weights scale
as $w_\bulk \approx 1$, $w_\edge \propto (\omega L)^{-1}$, and
$w_\vertex \propto (\omega L)^{-2}$. Jammed configurations become
negligible, particles are almost always free, distributed
quasi-uniformly over the torus. This is consistent with the now
frequent velocity tumbles leading to a motion that is essentially
diffusive.

These two asymptotic regimes already appear in 2-RTP systems, as shown
for both the steady state and mixing time~\cite{hahn23,guillin24}. The
striking novelty in the 3-RTP case is the emergence of a
\emph{separated} regime at finite $\omega L=O(1)$, in which two
particles are jammed while the third remains free, leading to a
slowdown of the average velocity.  Here, the separated phase dominates
the invariant measure: the most probable configurations consist of two
particles jamming while the third remains free. The sharp growth of
$w_\edge$ along $\omega L$ and its slow subsequent decay highlight the
robustness of this regime. The secondary role of the relaxation
component in the invariant measure is now apparent. For all values of
$\omega L$, the relaxation contribution $w_\edge w_\rel$ always
remains subdominant in comparison to the structural ones
$w_\bulk,w_\edge w_\unif$ and $w_\vertex$. Moreover, in the diffusive
regime, relaxation, controlled by the scale $\lambda$, is
essentially instantaneous and the weight $w_\rel$ scales as
$\sim (\omega L)^{-1}$. In the persistent regime, $w_\rel\sim 1/3$ but
$\gamma$ is effectively uniform.  Thus, relaxation plays at most a
secondary role relative to the dominant structural weights, primarily
mediating the impedance between boundaries. Its limited impact, also
observed in the general 2-RTP case~\cite[Section VI]{hahn23},
indicates that relaxation---sometimes reinterpreted as an effective
attraction---cannot be regarded as the primary driver of
clustering. Even in $N$-particle systems, the formation and stability
of clusters are therefore expected to be dictated by the dominant
structural weights.\smallskip

\section*{N-RTP steady state}

\begin{figure}
  \includegraphics[width=0.95\columnwidth]{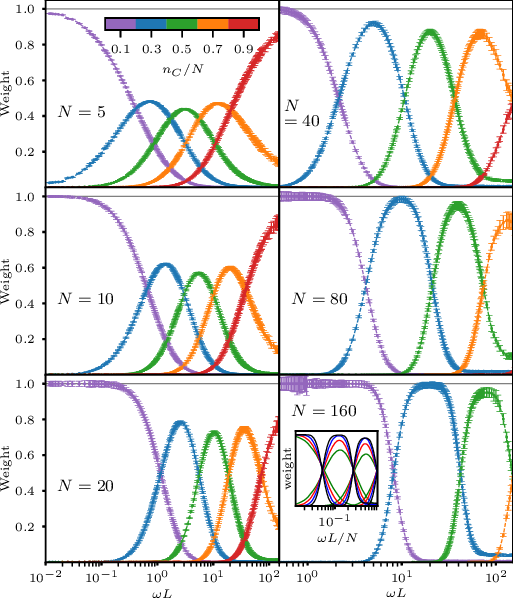}
  \caption{Probability weights to observe $n_C$ clusters as a function
    of $\omega L$ for different system sizes $N$ (simulation details
    in the Appendix~\ref{sec:simu}). Each curve gathers configurations with
    $n_c/N$ within intervals $[0.2\, m,0.2\, (m+1)]$. A progressive
    sharpening of the crossovers between $n_c$ intervals is observed
    as $N$ increases. Inset: Same data for $N=20$ (green), $40$ (red),
    $80$ (blue), and $160$ (black), with $\omega L$ rescaled by $N$;
    the curves point at an asymptotic collapse onto a single master
    curve.  }
        \label{fig:N-RTP}
\end{figure}

A central open question concerns the fate of the separated regime in
the $N$-particle case. Understanding the nature of clustering in this
regime, i.e.~whether a single macroscopic cluster forms, or the system
fragments into multiple clusters, or displays fractal behavior,
remains an important open question. While a full solution of the
$N$-particle steady state lies beyond the scope of this work, the
three-RTP case already reveals structural features that extend
naturally to general $N$. For instance, the persistent and diffusive
regimes extrapolate directly: for $\omega L \gg 1$ the process is
expected to approach a uniform distribution on the simplex, whereas
for $\omega L \ll 1$ the probability mass concentrates on subsets of
jammed configurations (no free particles). More importantly, the
essential new insights arise in the non-asymptotic regime:

\noindent (i) The building blocks of the three-body steady state
generalizes straightforwardly to the N-body steady state $\pi_N$:
$(n_C\!-\!1)$-dimensional faces associated with configurations
exhibiting $n_C$ clusters ($1\leq n_C\leq N$), where a free particle
is treated as a cluster of size one. The $(N-1)$-dimensional face thus
corresponds to the bulk region with all particles being free
($n_C = N$). A hierarchy of probability flows between these faces sets 
their relative weights $w_{n_C}$,
\begin{equation}
  \pi_N = \sum_{n_C=1}^Nw_{n_C}\pi^{(n_C)}_N,
\end{equation}
where $\pi^{(n_C)}_N$ is the probability distribution of
  configurations with $n_C$ clusters and decomposes in
  uniform and relaxation terms, as set by first-order PDE.

\noindent
(ii) The hierarchical scaling
of structural weights in powers of $\omega L$ depending on the number
of jamming particles persists for general $N$, as it
originates from the boundary exchange processes between run and tumble
terms. This produces a geometrical hierarchy of scales for the $w_{n_C}$'s,
\begin{equation}
  w_{n_C} = \frac{(\omega L)^{n_C} a_{n_C}(\omega L)}{\sum_l w_l}.
\end{equation}
The dependence in $\omega L$ of the $a_{n_C}$'s stems from relaxation
terms. Based on this relaxational origin and on the two- and
three-body steady states, we conjecture that the $a_{n_C}$ vary weakly
with $\omega L$, particularly away from the persistent regime.

\noindent (iii) This geometric hierarchy implies an exponential
divergence, for increasing $N$, of the slope of the cumulative
probability of observing a number of clusters $n_C$ proportional to system
size, namely
  $w_{n_C} \propto a_{n_C}\exp(\ln (\omega L)n_C)$.
Our numerical results support this prediction: Fig.~\ref{fig:N-RTP}
shows a progressive sharpening of the crossovers between
cluster-number regimes as $N$ increases. The simulations further
confirm a natural collapse of the curves when rescaling $\omega L$ by
$N$, consistent with a renormalization by the mean free path scale $L/N$. In
the thermodynamic limit, the steady state is therefore expected to
become a mixture over Dirac measures, such that for each value of the
renormalized activity parameter $\omega L/N$ the system exhibits a
well-defined number of clusters.

\noindent (iv) Accordingly, the activity $\ln(\omega L)$ can be
  interpreted as the fugacity conjugate to cluster creation, naturally
  leading to a grand-canonical description of the cluster number. This
  is consistent with clusters behaving as effective particles. Drawing
  an analogy with the equilibrium grand-canonical weight $\exp(-\beta E + \mu N)$,
  the term $\pi^{(n_C)}_N(r,\sigma)$ can be viewed as the effective
  interaction energy contribution, although it is not generally of
  product form along $r$ and $\sigma$, even if the uniform
  contributions are expected to dominate. However, the picture
  must be refined, since the prefactors $a_{k}$ also contribute to the
  weight hierarchy and stems from the nonequilibrium nature of
  the underlying dynamics. Indeed, the resulting
  “grand-canonical” structure does not arise from reversible exchanges
  with a reservoir, but from the intrinsically irreversible creation
  and annihilation of clusters induced by run-and-tumble
  persistence. Thus, while the effective-particle viewpoint makes the
  fugacity analogy natural, its origin is genuinely nonequilibrium and
  subtler than a direct mapping onto the thermodynamical
  grand-canonical ensemble.

\section*{Discussion}

Our results provide an exact microscopic characterization of
activity-driven clustering beyond two particles. By resolving the
three-RTP steady state, we revealed that nontrivial many-body
configurations emerge purely from the interplay of run-and-tumble
dynamics and hard-core repulsion. In particular, the separation of
time scales induced by jamming generates distinct regimes (persistent,
separated, and diffusive) controlled solely by the activity parameter
$\omega L$. Remarkably, the separated phase, in which two particles
are jammed while the third remains free, dominates at finite
activity. Furthermore, the relaxation contributions, often interpreted as
mediators of attraction, are in fact subdominant, confirming that
structural weights primarily govern the system's organization.

This analysis is made possible by the PDMP approach we develop. It
offers an alternative and complementary perspective to the
effective-attraction picture often used in coarse-grained
theories~\cite{tailleur08,fily12,cates15}. It is particularly valuable
for models with jamming in the absence of thermal noise, where exact
hydrodynamic limits remain elusive~\cite{erignoux21, kourbane18}. The
key framework here is that of PDMPs, whose generator provides an
efficient mean to handle the otherwise delicate boundary conditions in
the Fokker-Planck formulation~\cite{hahn23}.  The invariance condition
then translates into a hierarchy of PDEs, encoding the conservation of
probability flows under deterministic RTP motion and stochastic
velocity flips.

While the full $N$-particle steady state is beyond the scope of this
work, we were able to cut through its combinatorial complexity and
extract its essential structure. Extending the three-body insights, we
find that the hierarchical organization of clusters persists:
configurations with $n_C$ clusters naturally define
$(n_C-1)$-dimensional faces of the state space, and the associated
structural weights scale geometrically with $\omega L$. This geometric
hierarchy produces a sharply defined number of clusters in the
thermodynamic limit, with the slope of cumulative cluster
probabilities diverging as $N$ grows. Viewed through the lens of
statistical mechanics, the activity parameter closely acts as a
fugacity conjugate to cluster number, providing a nonequilibrium
analogue of a grand-canonical ensemble. Crucially, this
effective-particle interpretation emerges directly from microscopic
dynamics rather than coarse-grained assumptions, bridging a key gap
between local rules and macroscopic steady-state organization.

A crucial open question is how the competing geometric scales in
  the weight hierarchy more precisely interact and whether a precise
  configuration at fixed cluster number dominates at a given
  $\omega L$. This issue lies at the heart of achieving a full
  microscopic characterization of the resulting cluster-size
  distribution, numerically shown as exponential
  \cite{soto14,sepulveda16}, determining whether separated phases
  survive for large $N$ and, more broadly, establishing whether MIPS
  can be given a rigorous microscopic characterization, or perhaps
  ruled out, via exact invariant-measure arguments. Our results
  highlight how delicate this balance is: clustering outcomes depend
  sensitively on activity strength, suggesting that a full resolution
  may ultimately require the complete combinatorial solution of the
  $N$-RTP steady state.

Ultimately, these results
  provide a principled microscopic foundation for understanding
  many-body steady states in active systems. They reveal how
  persistent motion and jamming alone suffice to stabilize phase
  separation and clustering. Beyond their fundamental significance,
  the developed framework and results open the door to predictive
  control of cluster formation in experiments and simulations, as the
  activity parameter provides a tunable handle to select desired
  cluster numbers and structures.

\smallskip\begin{acknowledgments}All the authors acknowledge the support of the
  French ANR under the grant ANR-20-CE46-0007 (\emph{SuSa} project).
  LH~acknowledges support from the Swiss National Science Foundation
  (grant no.~200029-21991311). A.G. has benefited from the support of a government grant managed by the Agence Nationale de la Recherche under the France 2030 investment plan ANR-23-EXMA-0001.
\end{acknowledgments}

\bibliography{biblio}
\clearpage

  \appendix
  \onecolumngrid
\section{3-RTP steady state}
\label{app:full-pi}
We summarize here the complete expression of the 3-RTP steady state, for $(r,\sigma)\in \Omega$,
\begin{equation}
  \pi(r, \sigma) = w_{\mathcal{B}}\frac{\1_\bulk(r,\sigma)}{4 L^2}  + \frac 13 \sum\limits_{k=1}^3\Big(w_\edge \1_{\edge_k}(r,\sigma)\left[w_\mathrm{eq}\frac{1+\delta_{\sigma_1,-\sigma_2}}{8L} + w_\mathrm{rel} \gamma(r_{k+1},\tau_k(\sigma))\right]
  + w_{\vertex} \1_{\vertex_k}(r,\sigma)\frac{1+2\delta_{\sigma_1,-\sigma_3}}{8}
\Big),
\label{eq:full-pi}
\end{equation}
with,
\begin{equation*}
  \gamma(\bar{r}, \sigma) = \frac{\omega\sqrt 2(2 + (\sigma_1\sigma_3-2)\delta_{\sigma_1,\sigma_2}) }{8\text{sh}(\omega L \sqrt 2)} \Big[(2-\delta_{\sigma_1,\sigma_2}) \text{ch}\left(2\sqrt 2 \omega(\tfrac{L}{2} - \bar{r})\right) + \sigma_{3}\sqrt 2 \text{sh}\left(2\sqrt 2 \omega(\tfrac{L}{2}-\bar{r})\right)\Big],
\end{equation*}
and,
\begin{align*}
  w_\unif&= \frac{ \omega L\left[\frac{2\sqrt 2}{{\rm th}( \omega L\sqrt 2)}  + \frac{10}{3}\right]}{1+ \omega L\left[\frac{2\sqrt 2}{{\rm th}( \omega L\sqrt 2)}  + \frac{10}{3}\right]},&
      w_{\text{rel}}&=  \frac{1}{1+ \omega L\left[\frac{2\sqrt 2}{{\rm th}( \omega L\sqrt 2)}  + \frac{10}{3}\right]},&\\
  w_\bulk&=\frac{(\omega L)^2w_\unif}{2\omega L+ w_\unif\left[6+4\omega L + (\omega L)^2 \right]}, &
   w_\edge&=\frac{6\omega L}{2\omega L+ w_\unif\left[6+4\omega L + (\omega L)^2 \right]}, 
 w_\vertex&=\frac{ ( 6+4\omega L)w_\unif -4\omega L}{2\omega L+ w_\unif\left[6+4\omega L + (\omega L)^2 \right]}.
\end{align*}
The partition of the state space $\Omega$ into $\bulk$ (free states), $\cup_k\edge_k$ (separated states) and
$\cup_k\vertex_k$ (jammed states) is precisely defined in Appendix~\ref{app:state}.
\section{State space definition}
\label{app:state}
In this section, we provide a detailed and precise definition of the
state space, as this is essential for computing the integral in the
invariance condition. The subtlety lies in the integration by parts,
where the boundary terms must be specified with great care in order to
determine unambiguously which states belong to which set. A state is
assigned to a given set (separated or free) only if it does not
immediately leave it under the dynamics (i.e., through the creation or
annihilation of a jamming pair). This makes a careful understanding of
the dynamics indispensable, and we therefore propose here an explicit
construction of the state space, illustrated in
Fig.~\ref{fig:main_figure}.\medskip

The state space for $r$ is split into:
\begin{itemize}
  \item the simplex
    $\Delta = \{ r \in ]0,L[^3; \sum_k r_k = L \}$
  \item the three edges
    $\partial_k \Delta =\{r;\tau_k(r)\in \{0\}\times ]0,L[^2 \
    \text{and} \ \sum_k r_k = L\}$ with normal
    $n^\edge_k = -( \delta_{1,k}, \delta_{2,k},\delta_{3,k})$
  \item the three vertices
    $\partial^2_{k} \Delta=\{r; \tau_k(r)= (0,0,L)\}$ with normal
    $n^\vertex_{k} = ( \delta_{1,(k+2)},
    \delta_{2,(k+2)},\delta_{3,(k+2)})$.
\end{itemize}
The state space for the velocities is $\Sigma=\{-1,1\}^3$ and we
define the subsets
$\Sigma_{k^\pm}=\{\sigma; \pm(\sigma_{k+1}-\sigma_k)>0\}$ and
$\Sigma_{k^0}=\{\sigma; \sigma_{k+1}=\sigma_k\}$. Using the
definition of the flow $\phi$ \eqref{eq:phi}, the complete state space $\Omega$ for $(r,\sigma)$
identifies with
$\mathcal{B}\cup(\cup_{k=1}^3\edge_k)\cup(\cup_{k=1}^3\vertex_k)$
where:
    \begin{itemize}
\item {(\it Free}) the bulk $\bulk$ identifies with the
three RTP being free, i.e. \begin{equation}\bulk = (\Delta \times \Sigma)\cup(\cup_k\partial_k{\bulk}^+),\end{equation} with
the sets of entering states
\begin{equation}\partial_k \bulk^+= \partial_k \Delta \times \Sigma_{k^+}.\end{equation}
The set of exit states (not included in $\bulk$) are
\begin{equation}\partial_k\bulk^-= \partial_k \Delta \times \Sigma_{k^-}.\end{equation}
The set $\cup_k\partial_k\bulk^+\cup_k\partial_k\bulk^-$ amounts to $\partial \bulk$ as source term in \eqref{eq:cs}.
\item {(\it Separated}) the edge $\edge_k$ correspond to RTP
  $k$ and $k+1$ jamming and the $k+2$-one being free, i.e.
  \begin{equation}\edge_k= (\partial_k \Delta \times
    (\Sigma\setminus\Sigma_{k^+}))\cup\partial_k{\edge}_{k}^{+}\cup\partial_{k+2}{\edge}_{k}^{+}\end{equation}
  with  the set of entering states into $\edge_k$ from vertex $\vertex_{k}$ and $\vertex_{k+2}$, 
  \begin{equation}\partial_k\edge_{k}^{+} = \partial^2_{k} \Delta\times(\Sigma_{(k+1)^+}\setminus \Sigma_{k^+}),
  \end{equation}
  and,
     \begin{equation}
    \partial_{k+2}\edge_{k}^{+}=\partial^2_{k+2} \Delta\times(\Sigma_{(k+2)^+}\setminus\Sigma_{k^+}).\end{equation}
  It includes exit states from $\bulk$.
  The sets of exit states (not included in $\edge_k$) are
  \begin{equation}\partial_k{\edge}_{k}^-=  \partial^2_{k} \Delta\times(\Sigma_{(k+1)^-}\setminus \Sigma_{k^+}),\end{equation}
 and, \begin{equation}
    \partial_{k+2}{\edge}_{k}^-= \partial^2_{k+2} \Delta\times(\Sigma_{(k+2)^-}\setminus\Sigma_{k^+}).\end{equation}
  The set $\partial_{k'}\edge_k^+\cup\partial_{k'}\edge_k^-$ amounts to $\partial_k\edge_{k'}$ as source term in \eqref{eq:cj} (note that $\partial_{k+1}\edge_k=\emptyset$).
\item {(\it Jamming}) the vertex
  \begin{equation}\vertex_k = \partial^2_{k}\Delta \times
    (\Sigma\setminus(\Sigma_{k^+}\cup\Sigma_{(k+1)^+}))\end{equation} corresponds
  to the three RTP jammed in the order $k,k+1,k+2$. It includes
  exit states from $\edge_k$ and $\edge_{k+1}$.
\end{itemize}    
\section{Conservation constraints}\label{app:conservation}
Plugging the generator expression \eqref{eq:generator} into the
invariance condition $\int_{\Omega} \mathcal Af\, d\pi = 0$, we obtain
for $f$ a continuous test function:
\begin{equation}
    \int_{\bulk} \langle  \phi(z), \nabla_r f(z) \rangle d\pi(z) +  \sum_k\int_{\edge_k} \langle  \phi(z), \nabla_r f(z) \rangle  d\pi(z)+ \int_{\bulk\cup(\cup_k\edge_k)\cup(\cup_k\vertex_k)}\omega\sum_{j}\left(f(\iota_j(z)) - f(z)\right) d\pi(z) = 0
\end{equation}
The first terms are treated by integration by parts, leading to,
\begin{equation}
   \int_{\bulk} \langle  \phi(z), \nabla_r f(z) \rangle d\pi(z) = -\int_{\bulk} \langle  \phi(z), \nabla_r \pi(z) \rangle f(z)dz + \sum_k\int_{\partial_k\bulk^+\cup\partial_k\bulk^-} \langle  \phi(z), n^\edge_k\rangle \lim_{\zeta\in\bulk\to z}\pi(\zeta) f(z)dz,
  \end{equation}
and,
\begin{multline}
  \int_{\edge_k} \langle  \phi(z), \nabla_r f(z) \rangle d\pi(z) = -\int_{\edge_k} \langle  \phi(z), \nabla_r \pi(z) \rangle f(z)dz + \sum_k\int_{\partial_{k}\edge^+_{k}\cup\partial_{k}\edge^-_{k}} \langle  \phi(z), n^\vertex_{k}\rangle  \lim_{\zeta\in\edge_k\to z}\pi(\zeta) f(z)dz\\
 + \sum_k\int_{\partial_{k+2}\edge^+_{k}\cup\partial_{k+2}\edge^-_{k}} \langle  \phi(z), n^\vertex_{k+2}\rangle \lim_{\zeta\in\edge_{k}\to z}\pi(\zeta) f(z)dz
\end{multline}
Since $f$ is an arbitrary test function, the relation must hold for
each value of $z$. By collecting all contributions associated with a
given $z$ and replacing the normals by their expression and $\pi$ by
its decomposition \eqref{eq:pi-gen}, we therefore arrive at the
conservation conditions \eqref{eq:cb},\eqref{eq:cs} and \eqref{eq:cj}.

\section{Run–Tumble flow diagrams and jamming–unjamming (im)balance at phase boundaries}
\label{app:tumble-fig}

\begin{figure}
  \centering
\begin{tikzpicture}[
  blacknode0/.style={rectangle, draw=black,  inner sep=0.3pt, line width=1pt}, 
  blacknode1/.style={rectangle, draw=black,  fill=black!15,  inner sep=0.3pt, line width=1pt}, 
  blacknode2/.style={rectangle, draw=black,  fill=black!50,  inner sep=0.3pt, line width=1pt}, 
  rednode/.style={rectangle, draw=red,   inner sep=0.3pt, line width=1pt}, 
  greennode/.style={rectangle, draw=green!60!black!80,  inner sep=0.3pt, line width=1pt}, 
    arrow/.style={-Stealth, line width=1pt}]

\node[blacknode1] (pmm) at (0,-1.5) {$\parrowf\marrowf\ \  \marrowf$};
\node[blacknode1] (pmp) at (0,0) {$\parrowf\marrowf\ \  \parrowf$};

\draw[arrow] (pmm)  to[bend left=20] node[midway,left] {$\omega$} (pmp);
\draw[arrow] (pmp)  to[bend left=20] node[midway,right] {$\omega$} (pmm);

\node[blacknode0] (ppp) at (3.5,1) {$\parrowf\parrowf\ \ \parrowf$};
\node[blacknode0] (mmm) at (3.5,-2.5) {$\marrowf\marrowf\ \ \marrowf$};
\node[blacknode2] (ppm) at (2.8,-1.5) {$\parrowf\parrowf\ \ \marrowf$};
\node[blacknode2] (mmp) at (2.8,0) {${\marrowf\marrowf}\ \ \parrowf$};

\draw[arrow] (ppp.340) to[in=0, out=-40] node[pos=0.18,right]{$\omega$} (ppm.1);
\draw[arrow] (ppm.15) to[out=0, in=-40] node[pos=0.79, left] {$\omega$} (ppp.335);
\draw[arrow] (mmm.20) to[in=0, out=40] node[pos=0.18,right] {$\omega$} (mmp.350);
\draw[arrow] (mmp.345) to[out=0, in=40] node[pos=0.78,left] {$\omega$} (mmm.25);

\draw[arrow] (mmm.175) to[out=180, in=-60] node[midway,above] {$\omega$} (pmm.280);
\draw[arrow] (pmm.260) to[out=-90, in=180] node[midway,below] {$\omega$} (mmm.185);
\draw[arrow] (ppp.185) to[out=180, in=60] node[midway,below] {$\omega$} (pmp.80);
\draw[arrow] (pmp.100) to[out=90, in=180] node[midway,above] {$\omega$} (ppp.175);

\draw[arrow] (ppm) to[bend left=5] node[midway,below] {$\omega$} (pmm);
\draw[arrow] (pmm) to[bend left=5] node[midway,above] {$\omega$} (ppm);
\draw[arrow] (mmp) to[bend left=5] node[midway,below] {$\omega$} (pmp);
\draw[arrow] (pmp) to[bend left=5] node[midway,above] {$\omega$} (mmp);

\node[rednode] (ipmp) at (-2.8,0) {$\textcolor{black}{\parrowf} \marrowf \ \ \parrowf$};
\node[rednode] (ipmm) at (-2.8,-1.5) {$\textcolor{black}{\parrowf} \marrowf \ \ \textcolor{black}{\marrowf}$};
\draw[arrow,red] (ipmm) --  node[midway,below] {$\phi$}  (pmm);
\draw[arrow,red] (ipmp) --  node[midway,above] {$\phi$}  (pmp);

\node[greennode] (ompp) at (7.5, 0.0) {$\textcolor{black}{\marrowf}\parrowf\  \ \textcolor{black}{\parrowf}$};
\node[greennode] (ompm) at (7.5,-1.5) {$\textcolor{black}{\marrowf}\parrowf\ \ \marrowf$};
\draw[arrow,green!60!black!80] (ppp) --  node[pos=0.7,above,sloped] {$\omega\to\phi$} (ompp);
\draw[arrow,green!60!black!80] (mmp) -- node[pos=0.7,below] {$\omega\to \phi$} (ompp);
\draw[arrow,green!60!black!80] (ppm) -- node[pos=0.7,above] {$\omega \to \phi$} (ompm);
\draw[arrow,green!60!black!80] (mmm) --  node[pos=0.7,below,sloped] {$\omega\to\phi$} (ompm);

\end{tikzpicture}
\caption{Diagram of configuration relationships at fixed $r$
  within the separated phase and at the separated–free boundary. Black
  contours denote separated states, red indicate configurations
  becoming jammed, and green mark unjamming states. Grey shading
encodes the relative speed of the free particle in the separated phase
(white for speed 0, light grey for speed 1, dark grey for speed 2).}
\label{fig:diag-sep}
\end{figure}

\begin{figure}
  \centering
\begin{tikzpicture}[
  blacknode/.style={rectangle, draw=black, inner sep=0.3pt, line width=1pt}, 
  rednode1/.style={rectangle, draw=red, fill=black!15,  inner sep=0.3pt, line width=1pt}, 
  rednode2/.style={rectangle, draw=red,  fill=black!50,  inner sep=0.3pt, line width=1pt}, 
  greennode1/.style={rectangle, draw=green!60!black!80, fill=black!15, inner sep=0.3pt, line width=1pt}, 
  greennode2/.style={rectangle, draw=green!60!black!80, fill=black!50, inner sep=0.3pt, line width=1pt}, 
    arrow/.style={-Stealth, line width=1pt}]

\node[blacknode] (ppp) at (2.5,0.) {$\parrowf\parrowf\parrowf$};
\node[blacknode] (mmm) at (2.5,-2) {$\marrowf\marrowf\marrowf$};
\node[blacknode] (pmm) at (0,-2) {$\parrowf\marrowf\marrowf$};
\node[blacknode] (ppm) at (0,0) {$\parrowf\parrowf\marrowf$};

\draw[arrow] (pmm)  to[bend left=20] node[midway,left] {$\omega$} (ppm);
\draw[arrow] (ppm)  to[bend left=20] node[midway,right] {$\omega$} (pmm);

\draw[arrow] (ppp) to[bend left=5] node[midway,below] {$\omega$} (ppm);
\draw[arrow] (ppm) to[bend left=5] node[midway,above] {$\omega$} (ppp);
\draw[arrow] (mmm) to[bend left=5] node[midway,below] {$\omega$} (pmm);
\draw[arrow] (pmm) to[bend left=5] node[midway,above] {$\omega$} (mmm);

\node[rednode2] (i1pmm) at (-3,-2.5) {$\textcolor{blue}{\parrowf}\ \textcolor{black}{\marrowf\marrowf}$};
\node[rednode1] (i1ppm) at (-3,-0.5) {$\textcolor{blue}{\parrowf}\ \textcolor{black}{\parrowf\marrowf}$};
\node[rednode1] (i2pmm) at (-3,-1.5) {$\textcolor{black}{\parrowf\marrowf} \ \textcolor{blue}{\marrowf}$};
\node[rednode2] (i2ppm) at (-3,0.5) {$\textcolor{black}{\parrowf\parrowf}\ \textcolor{blue}{\marrowf}$};
\draw[arrow,red] (i1pmm) --  node[midway,below] {$\phi\!=\!1$}  (pmm);
\draw[arrow,red] (i2pmm) --  node[midway,above] {$\phi\!=\!2$}  (pmm);
\draw[arrow,red] (i1ppm) --  node[midway,below] {$\phi\!=\!1$}  (ppm);
\draw[arrow,red] (i2ppm) --  node[midway,above] {$\phi\!=\!2$}  (ppm);

\node[greennode1] (opmp) at (6.5, -0.5) {$\textcolor{black}{\parrowf\marrowf}\ \textcolor{blue}{\parrowf}$};
\node[greennode2] (ompp) at (6.5,0.5) {$\textcolor{blue}{\marrowf} \ \textcolor{black}{\parrowf\parrowf}$};
\node[greennode1] (ompm) at (6.5,-1.5) {$\textcolor{blue}{\marrowf}\ \textcolor{black}{\parrowf\marrowf}$};
\node[greennode2] (ommp) at (6.5,-2.5) {$\textcolor{black}{\marrowf\marrowf}\ \textcolor{blue}{\parrowf}$};
\draw[arrow,green!60!black!80] (pmm) to[out=30,in=180] node[pos=0.44,above] {$\omega\to\phi=1$}  (opmp.south west);
\draw[arrow,green!60!black!80] (ppp) --  node[midway,below,sloped] {$\omega\to\phi=1$}  (opmp);
\draw[arrow,green!60!black!80] (ppp) --   node[midway,above,sloped] {$\omega\to\phi=2$} (ompp);
\draw[arrow,green!60!black!80] (ppm) to[out=-30, in=-180] node[pos=0.44,below] {$\omega\to \phi=1$} (ompm.north west);
\draw[arrow,green!60!black!80] (mmm) --  node[midway,above,sloped] {$\omega\to\phi=1$} (ompm);
\draw[arrow,green!60!black!80] (mmm) --  node[midway,below,sloped] {$\omega\to\phi=2$} (ommp);

\end{tikzpicture}
\caption{Diagram of configuration relationships within the jammed
  phase and at the jammed–separated boundary. Configurations with
  black contours correspond to jammed states, red contours indicate
  states becoming jammed, and green contours represent unjamming
  states. The grey shading denotes the relative speed of the jamming
  or unjamming free particle (light grey for speed
  1, dark grey for speed 2).}
\label{fig:diag-jam}
\end{figure}

We provide diagrams of the flow exchanges in the
separated (Fig.~\ref{fig:diag-sep}) and jammed
(Fig.~\ref{fig:diag-jam}) phases. They show that detailed jamming is
trivially satisfied at the free–separated boundary
($P(\textcolor{red}{\parrow\marrow}\parrow)=P(\textcolor{green!60!black!80}{\marrow\parrow}\parrow)$)
but necessarily broken at the separated–jammed one. A direct
inspection makes this evident: enforcing detailed balance on the
jamming/unjamming flows is impossible, since the tumble steady states
of the two phases do not match. Specifically, the separated tumble
steady state requires
\[P(\textcolor{red}{\parrow} \parrow\marrow)=2P(\parrow\parrow\textcolor{red}{\marrow})\]
while the jammed steady state enforces
\[P(\textcolor{green!60!black!80}{\marrow}\parrow\marrow)=8P(\marrow\marrow\textcolor{green!60!black!80}{\parrow})\]
demonstrating the incompatibility and the breaking of the
detailed-jamming symmetry. As a consequence, the solution in the
separated phase can no longer reduce to a uniform form proportional to
the tumble steady state; an additional relaxation contribution is
required to reconcile the imbalance.

\section{Relaxation equation in the separated phase}\label{app:relax}
By symmetry, the distribution $\pi_{\edge_k}$ simplifies without loss
of generality to its expression in \eqref{eq:pi}. We determine
$\gamma$ by considering the case over $\edge_1$. We introduce the
vector
$\pi^\edge(\bar{r}) =
(\pi_{\edge_1}((0,\bar{r},L-\bar{r}),\sigma)_{\sigma \in \Sigma}$,
which contains all the densities on the edge
$\mathcal S_1 = \mathcal J_3 \mathcal J_1$. The parameter $\bar{r}$
lives in $[0, L]$ and $\bar{r} = 0$ corresponds to
$\mathcal J_1 = (0, 0, L)$, while $\bar{r} = L$ corresponds to
$\mathcal J_3 = (0, L, 0)$. Expliciting the condition \eqref{eq:cs}
following the diagram in Fig.~\ref{fig:diag-sep} leads to,
 \begin{align}
 \label{C+-+}\tag{$\mathrm{C}^\edge_{+-+}$} \partial_{\bar{r}}\pi^\edge_{+-+} &=   -3\omega\pi^\edge_{+-+} +\omega\pi^\edge_{+++} +\omega\pi^\edge_{--+} +\omega\pi^\edge_{+--} +K,\\
 \label{C+--}\tag{$\mathrm{C}^\edge_{+--}$} \partial_{\bar{r}}\pi^\edge_{+--} &=  3\omega\pi^\edge_{+--} -\omega\pi^\edge_{---} -\omega\pi^\edge_{+-+} -\omega\pi^\edge_{++-} -K,\\
 \label{C--+}\tag{$\mathrm{C}^\edge_{--+}$} 2\partial_{\bar{r}}\pi^\edge_{--+} &=   -3\omega\pi^\edge_{--+} +\omega\pi^\edge_{---} +\omega\pi^\edge_{+-+}, \\
 \label{C++-}\tag{$\mathrm{C}^\edge_{++-}$} 2\partial_{\bar{r}}\pi^\edge_{++-} &=  3\omega\pi^\edge_{++-} -\omega\pi^\edge_{+++} -\omega\pi^\edge_{+--}, \\
 \label{C-++}\tag{$\mathrm{C}^\edge_{-++}$} K &= \omega\pi^\edge_{+++} +\omega\pi^\edge_{--+},\\
 \label{C-+-}\tag{$\mathrm{C}^\edge_{-+-}$} K &= \omega\pi^\edge_{---} +\omega\pi^\edge_{++-},\\
 \label{C+++}\tag{$\mathrm{C}^\edge_{+++}$} 3\omega\pi^\edge_{+++} &= \omega\pi^\edge_{++-} +\omega\pi^\edge_{+-+},\\
 \label{C---}\tag{$\mathrm{C}^\edge_{---}$} 3\omega\pi^\edge_{---}  &=  \omega\pi^\edge_{--+} +\omega\pi^\edge_{+--}.
 \end{align}
where $$K=\frac{3w_\bulk}{w_\edge}\frac{1}{2L^2}$$ stems for the source
terms from the bulk (that are constant as we restrict possible
solutions to being the uniform distribution $\frac{2}{L^2}\mu_\bulk$
in $\bulk$ and $\mu_\bulk(\sigma)=1/8$).  Using \eqref{C-++},
\eqref{C-+-}, \eqref{C+++} and \eqref{C---}, we reduce the system of
equations to the following systems on states with nonzero flow $\phi$,
\begin{align}
  \label{eq:first-order}
  \partial_{\bar r}\left(\begin{array}{@{}l@{}}
    \pi^\edge_{+-+}\\
    \pi^\edge_{--+}\\
    \pi^\edge_{+--}\\
    \pi^\edge_{++-}
\end{array}\right) &\!=  \!
                    \frac{\omega}{6} \left(\begin{array}{@{}cccc@{}}
    -14 & 12 & 6  & 4    \\
    3 & -8 & 1 & 0   \\
    -6& -4 &14  & -12   \\
    -1& 0 & -3 & 8   \\   
  \end{array}\right)                     
                     \left(\begin{array}{@{}l@{}}
                       \pi^\edge_{+-+}\\
                       \pi^\edge_{--+}\\
                       \pi^\edge_{+--}\\
                       \pi^\edge_{++-}
  \end{array}\right).
\end{align}
This resulting matrix coincides with the matrix $B$ in \cite{hahn23}
if using a mapping to an effective two-particle system with a tumble
mechanism modified by the bulk source terms.

It leads to the following second-order relaxation equations that recover  \eqref{eq:2ndPDE},
\begin{align}
  \label{eq:relax+}
  \partial_{\bar{r}}^2(\bm{\pi}^+_{\edge_k}+\bm{\pi}^-_{\edge_k}) &= 8\omega^2L_+(\bm{\pi}^+_{\edge_k}+\bm{\pi}^-_{\edge_k})
                       ,  L_+=\frac{1}{3}\left(
                       \begin{array}{@{}cc@{}}
                         2 &\!-\!4\\
                         \!-\!1/2 & 1
                       \end{array}\right)\\
                         \label{eq:relax-}
  \partial_{\bar{r}}^2(\bm{\pi}^+_{\edge_k}-\bm{\pi}^-_{\edge_k})
   &=8\omega^2 L_-(\bm{\pi}^+_{\edge_k}-\bm{\pi}^-_{\edge_k}),   L_-=\frac{1}{3} \left(\begin{array}{@{}cc@{}}
                                                 2 &-2\\
                                                 -1 & 1
                                               \end{array}\right),
\end{align}
where $\bm{\pi}^+_{\edge_1}=(\pi^\edge_{+-+}, \pi^\edge_{--+})^T $,
$\bm{\pi}^-_{\edge_1}=(\pi^\edge_{+--}, \pi^\edge_{++-})^T$, and more generally, $\bm{\pi}^\pm_{\edge_k}(\bar{r})=(\pi_{\edge_k}(r,\sigma^\pm_n))_{n=1}^2$,
with $r_{k+1}=\bar{r}$ and $\sigma^\pm_n$ so that the flow
$\phi_{k+1}(r,\sigma^\pm_n) = \pm n$.

The matrix $L_+$ admits $\bm{\mu}_\edge =(2,1)$ for eigenvector of
eigenvalue 0 and $\bm{\nu}_S=(4,-1)$ for eigenvector of eigenvalue
1. The matrix $L_-$ admits $\bm{\mu}_-=(1,1)$ for eigenvector of
eigenvalue 0 and $\bm{\nu}_A= \sqrt 2(2,-1)$ for eigenvector of
eigenvalue 1. The fact that $\bm{\mu}_\edge\not\propto\bm{\mu}_-$ is
consistent with the breaking of the detailed-jamming symmetry
\cite{hahn23}. The solution necessarily decomposes into,
\begin{align}
  \bm{\pi}^+_{\edge_k}(\bar{r})+ \bm{\pi}^-_{\edge_k}(\bar{r})= w_\unif\bm{\mu}_\edge + w_{\rm rel}\bm{\nu}_S\Big[&a_+\text{ch}\Big(\lambda\Big(\frac{L}{2}-\bar{r}\Big)\Big)
                                                                                                                    +b_+\text{sh}\Big(\lambda\Big(\frac{L}{2}-\bar{r}\Big)\Big)\Big] \label{eq:minus}\\
  \bm{\pi}^+_{\edge_k}(\bar{r})-  \bm{\pi}^-_{\edge_k}(\bar{r}) = w\bm{\mu}_- + w_{\rm rel}\bm{\nu}_A\Big[&a_-\text{ch}\Big(\lambda\Big(\frac{L}{2}-\bar{r}\Big)\Big)
                                                                                                            +b_-\text{sh}\Big(\lambda\Big(\frac{L}{2}-\bar{r}\Big)\Big)\Big]
                                                                  \label{eq:plus}
  \end{align}
  By symmetry, $ \bm{\pi}^+_{\edge_k}(L/2) =\bm{\pi}^-_{\edge_k}(L/2)$, therefore,
  \begin{equation}
    w=0, \; a_- = 0.
    \end{equation}
    Plugging \eqref{eq:plus} and \eqref{eq:minus} into \eqref{eq:first-order} yields,
    \begin{equation}
      a_+ = b_-, \; b_+ = 0,
      \end{equation}
      and deriving \eqref{C-++} and \eqref{C-+-} leads to
      $\partial\pi^\edge_{+++} = -\partial \pi^\edge_{--+}$ and
      $\partial\pi^\edge_{---} = -\partial \pi^\edge_{++-}$, recovering the
      expressions for $\nu_S$ and $\nu_A$ in \eqref{eq:nu}. The vector
      $\bm{\mu}_\edge$ identifies with the edge tumble measure
      $\mu_\edge$ in \eqref{eq:mus}.
  \section{Jammed constraints}
  \label{app:jam}

  By symmetry, the distribution $\pi_{\vertex_k}$ simplifies without
  loss of generality to its expression in \eqref{eq:pi} and can be
  determined by considering the case over $\vertex_1$. We introduce
  the vector $\mu = (\mu_\vertex(\sigma))_{\sigma \in \Sigma}$, which
  contains all the densities on the edge $\vertex_1$. Expliciting the
  condition \eqref{eq:cj} according to the diagram in Fig.~\ref{fig:diag-jam}
  leads to,
 \begin{align}
 \label{Cj+++}\tag{$\mathrm{C}^\vertex_{+++}$}  \omega\mu_{++-}=&3\omega\mu_{+++} ,\\
 \label{Cj---}\tag{$\mathrm{C}^\vertex_{---}$}   \omega\mu_{+--}=& 3\omega\mu_{---},\\
 \notag   \omega\mu_{---}+\omega\mu_{++-}=& 3\omega\mu_{+--}\label{Cj+--}\tag{$\mathrm{C}^\vertex_{+--}$}   -\frac{w_\edge}{w_\vertex}(\pi^\edge_{+--}(0^+)+2\pi^\edge_{--+}(L^-)) ,\\
  \notag \omega\mu_{+++}+\omega\mu_{+--}=&3\omega\mu_{++-} 
    \label{Cj++-}\tag{$\mathrm{C}^\vertex_{++-}$} -\frac{w_\edge}{w_\vertex}(\pi^\edge_{+-+}(L^-)+2\pi^\edge_{++-}(0^+)) , \\
 \label{Cj--+}\tag{$\mathrm{C}^\vertex_{--+}$} \omega\mu_{---}= & 2\frac{w_\edge}{w_\vertex}\pi^\edge_{--+}(0^+)  , \\
 \label{Cj-++}\tag{$\mathrm{C}^\vertex_{-++}$}\omega\mu_{+++} = &  2\frac{w_\edge}{w_\vertex}\pi^\edge_{++-}(L^-) ,\\
 \label{Cj-+-}\tag{$\mathrm{C}^\vertex_{-+-}$}  \omega\mu_{---}+\omega\mu_{++-}=&  \frac{w_\edge}{w_\vertex}\pi^\edge_{+--}(L^-) ,\\\label{Cj+-+}\tag{$\mathrm{C}^\vertex_{+-+}$}  \omega\mu_{+++}+\omega\mu_{+--}=& \frac{w_\edge}{w_\vertex}\pi^\edge_{+-+}(0^+),
 \end{align}
where the vector $\pi^\edge$ is defined in previous section~\ref{app:relax}.

Conditions \eqref{Cj+++} and \eqref{Cj---} and symmetry (or
equivalenty the equality of \eqref{Cj--+}/\eqref{Cj-++} or
\eqref{Cj-+-}/\eqref{Cj+-+}) lead to,
\begin{equation}
  \mu_\vertex((\pm 1,\pm 1, \pm 1)) =\frac{1}{8}, \quad  \mu_\vertex(( 1,\pm 1, - 1)) =\frac{3}{8}.
  \end{equation}
  Expliciting the vector $\mu$ and combining \eqref{Cj--+} and \eqref{Cj+-+} fix the ratio
  $w_\unif$ and $w_{\rm rel}$ and recovers \eqref{eq:muj},
\begin{equation}
  \begin{aligned}
    8\pi^\edge_{--+}(0^+)  &=\pi^\edge_{+-+}(0^+)\\
    \frac{3w_\unif}{4L}  &=   \frac{w_{\rm rel}}{\mathcal{N}_\gamma}\Big[12\text{ch}\Big(\lambda\frac{L}{2}\Big)+10\sqrt 2 \text{sh}\Big(\lambda\frac{L}{2}\Big)\Big]\\
    w_\unif  &=   w_{\rm rel}\omega L\left[\frac{2\sqrt 2}{\text{th}\Big(\omega L \sqrt 2\Big)}+\frac{10}{3}\right]
    \label{eq:weq-rel}
  \end{aligned}
  \end{equation}
  
  \section{Weight determination}
  \label{app:weights}
  
  We set the values of the weights
  $w_\bulk, \ w_\edge, \ w_\vertex$ through the conditions involving
  the boundary terms in \eqref{eq:cs} (e.g. \eqref{C-++}), 
  \begin{equation}
    3\frac{w_\bulk}{2L^2} = \frac{\omega}{4L}w_\edge w_\unif \to   w_\bulk = \frac{\omega L w_\unif}{6} w_\edge,
  \end{equation}
  and the boundary terms in \eqref{eq:cj} (e.g. \eqref{Cj+--}), 
  \begin{align}
    w_\edge\left(\frac{w_\unif}{2L} + w_{\rm rel}\omega  \frac{\sqrt 2}{4\text{th}\Big(\omega L \sqrt 2\Big)}\right)& = w_\vertex \frac{5\omega }{8}\\
     w_\unif(3+2\omega L) - 2\omega L & =3\omega L\frac{w_\vertex}{w_\edge},
    \end{align}
    and the normalisation condition,
    \begin{equation}
  w_\bulk + w_\edge + w_\vertex  = 1.
    \end{equation}
    We obtain,
    \begin{equation}
      \left(\frac{\omega L w_\unif}{6} + 1 + \frac{w_\unif(3+2\omega L) - 2\omega L}{3\omega L}\right)w_\edge = 1
    \end{equation}
    That fix the weights to,
    \begin{align}
      w_\bulk&= \frac{w_\unif(\omega L)^2}{2\omega L + w_\unif(6+4\omega L+ (\omega L)^2)}\\
      w_\edge &= \frac{6\omega L}{2\omega L + w_\unif(6+4\omega L+ (\omega L)^2)}\\
      w_\vertex&=\frac{ w_\unif(6+4\omega L) - 4\omega L}{2\omega L + w_\unif(6+4\omega L+ (\omega L)^2)}
      \end{align}

      \section{Simulation details}
\label{sec:simu}
      Simulations were performed by generating the PDMP corresponding
      to the considered RTP system, with $\omega =1$ and parameters
      $L, N$ set to the target values. Configurations were sampled
      every $\Delta t = 0.01$, yielding a total of $5.10^5$ recorded
      samples. For each parameter set, 50 (resp. 25) independent runs
      were performed for Fig.~\ref{fig:main_figure}
      (resp. Fig.~\ref{fig:N-RTP}). The error bars represent three
      times the standard deviation, computed as $\sigma/\sqrt{50}$
      (resp. $\sigma/\sqrt{25}$) where $\sigma$ is the standard
      deviation across each single run.

      As expected from \cite{guillin24}, the convergence time
      increases markedly at large $\omega L$, where the mixing time
      scales as $(\omega L)^2$ instead of linearly at smaller
      values. Consequently, the statistics in this regime are
      inherently noisier and require longer simulation runs to achieve
      comparable accuracy. While the steady state remain unchanged, we
      found that the choice of sampling interval $\Delta t$ also
      affects the statistical quality, e.g. too large a $\Delta t$ can
      delay the convergence of small-scale dynamical features at small
      $\omega L$ and reciprocally.
\end{document}